\documentclass[aps,pre,twocolumn,groupedaddress,showpacs]{revtex4-1}

\usepackage{bm,graphicx}
\usepackage{color}
\usepackage[dvips]{epsfig}
\begin{document}

\title{Patterns driven by combined AC and DC electric fields in nematic liquid crystals}

%Authors
\author{Alexei Krekhov}
\email[]{alexei.krekhov@ds.mpg.de}
\affiliation{Max Planck Institute for Dynamics and Self-Organization, D-37077 G\"ottingen, Germany}

\author{Werner Decker}
\author{Werner Pesch}
\affiliation{Physikalisches Institut, Universit\"at Bayreuth,
D-95440 Bayreuth, Germany}

\author{N\'{a}ndor \'{E}ber }
\author{P\'{e}ter Salamon }
\author{Bal\'{a}zs Fekete }
\author{\'{A}gnes Buka}
\affiliation{Institute for Solid State Physics and Optics,
Wigner Research Centre for Physics, Hungarian Academy of Sciences, H-1525 Budapest, P.O.B.~49, Hungary}

\date{\today}

%%% Abstract
\begin{abstract}
The effect of superimposed ac and dc electric fields on the formation of  electroconvection and flexoelectric patterns in nematic liquid crystals was studied.
For selected ac frequencies an extended standard model of the electro-hydrodynamic instabilities was used to characterize the onset of pattern formation in the two-dimensional parameter space of the magnitudes of the ac and dc electric field components.
Numerical as well as approximate analytical calculations demonstrate that depending on the type of patterns and on the ac frequency, the combined action of ac and dc fields may either enhance or suppress the formation of patterns.
The theoretical predictions are qualitatively confirmed by experiments in most cases.
Some discrepancies, however, seem to indicate the need to extend the theoretical description.
\end{abstract}

%\keywords{}

\maketitle

% Body of paper here - Use proper section commands

%
%%% Introduction
%
\section{\label{intro} Introduction}
A wide variety of pattern-forming instabilities in extended layers of nematic liquid crystals (nematics, NLCs) under the influence of electric field has been extensively investigated in the last decades (see, e.g., \cite{Buka:1996,Golovin:2006}).
Typically a NLC layer with thickness $d$ in the range $10$~$\mu$m~$ \lesssim d \lesssim 100$~$\mu$m is sandwiched between two electrode plates parallel to the $x, y$ plane to apply a voltage.
The confining plates are also used to ensure a homogeneous director orientation $\bm n \parallel \bm{\hat x}$ in the basic planar state, where $\bm n$ describes the locally preferred orientation of the nonspherical molecules of nematics.
The patterns that develop above a certain critical applied voltage are associated with spatially periodic director variations in the layer  plane, which are characterized by the critical wave vector $\bm q_c$.
Two different types of patterns have been observed in the past.
In some nematic materials and mostly under a dc electric field one finds the so-called flexodomains \cite{Barnik:1978,Blinov:1994}, where $\bm q_c \parallel \bm{\hat y}$.
Their analysis requires in essence only the minimization of the orientational free energy of nematics \cite{Bobylev:1977,Flexo:2011}.
In contrast to this equilibrium transition of the planar basic state, in the majority of cases the electroconvection instability (EC) is observed \cite{Gennes:1993,Pesch:1995}, i.e., a nonequilibrium transition accompanied by material flow, where the angle between $\bm q_c$ and $\bm{ \hat x}$ is quite small or often zero.
The theoretical analysis of the EC instability in nematics is based on the well established {\em standard model} (SM) \cite{Bodenschatz:1988} which has been further refined in the last two decades (see, e.g., \cite{Golovin:2006} and references therein).
In this paper we concentrate exclusively on the onset of pattern formation (linear regime).
However, the nonlinear regime is also well described by the standard model \cite{Plaut:1999,Plaut:1997}.
The basic ingredients of the SM (a coupled system of the Maxwell-, generalized Navier-Stokes- and director equations) are extensively discussed in the literature \cite{Gennes:1993,Blinov:1994}.
A key difference to isotropic fluids is that practically all material parameters are of tensorial nature and depend thus on $\bm n$, the distortions of which lead to charge separation (finite electric charge density $\rho_{el}$).
We deal only with nematics with a finite though very small electrical conductivity, which is due to certain mobile ions, often originating from the synthesis of the nematic compounds.
Note, however, that within the SM, nematics are simply modelled as anisotropic Ohmic conductors.
It should be emphasized that only the {\em internal} voltage drop seen by the nematic layer serves as input in the theory.
In principle it has to be distinguished from the {\em external} voltage applied to the confining plates.
In fact, the difficult task of relating these two voltages is not tackled within the present modeling of patterns in NLCs, in view of the complicated multilayer structure of the confining electrodes and the mostly unknown electrolytic properties of the NLCs.
Fortunately, in experiments, when (as in the majority of cases in the past) sinusoidal ac-voltages with not too small frequency, $f$, are used, the difference between the external and the internal voltage drop seems to be quite small, since the experiments match the theoretical calculations fairly well.
Here the ac-frequency $f$ (or the angular frequency $\omega=2\pi f$) serves as an important secondary control parameter besides the effective amplitude $U_0$ (the rms value) of the applied ac-voltage $U = \sqrt{2} U_0 \cos(\omega t)$.
It is useful, that for not too small $\omega$ the $\omega$-dependence of the main features of EC can often be absorbed by introducing the dimensionless frequency $\omega \tau_q$ with the charge relaxation time $\tau_q =\epsilon_0 \epsilon_{\perp}/\sigma_{\perp}$.
Here $\epsilon_{\perp}$ denotes the dielectric constant when an electric field is applied perpendicular to the director and $\sigma_{\perp} \sim 10^{-8} \, [\Omega$~m$]^{-1}$ is the corresponding small electric conductivity.
In contrast, the case of zero or very small $\omega$ is more complicated and challenging as well.
Consistent with the theory, a switching between flexodomains and EC patterns has been observed in some cases (see, e.g., \cite{bib:May-2008,Toth:2008,Eber:2012,Salamon:2013}).
In general the limit $\omega \rightarrow 0$ for EC is not smooth since the patterns flash up only during a very short fraction of the ac-period $T = 2 \pi/\omega$.
In addition large differences between the external and the internal voltage show up.
The linear properties of EC driven by a pure ac-field are described in detail on the basis of the SM in the literature.
In particular one is faced with two types of roll patterns of different time symmetry \cite{Nonsta:2008}.
For $\omega$ smaller than the so-called crossover frequency $\omega_c$ we have the {\em conductive} regime, where the time average of the out-of-plane director component, $n_z$, is finite in leading order and where the dimensionless wave number $\overline q_c=|\bm q_c| d/\pi$ of the pattern is of the order one.
For $\omega > \omega_c$ we have the {\em dielectric} regime where in  leading order $n_z \propto \cos(\omega t)$ and typically $\overline q_c \gtrsim 5$.
The existence of these two linear solution types can be traced back to a certain symmetry of the SM, which is invariant against the combination of a reflection at the midplane and a time-shift by  $T/2$.
Since the various convection patterns are associated with a periodically modulated director configuration in space, which has the effect of an optical grating (see, e.g., \cite{bib:Pesch-PRE-2013} and references therein), they are easily discriminated in experiments.
In the following sections we deal with the main topic of this work, namely the detailed description of the various pattern-forming instabilities in a nematic layer driven by two superimposed voltages, where each of them would separately trigger patterns of different time symmetry.
Section~\ref{theory} is devoted to the linear stability analysis of the underlying nemato-electrohydrodynamic equations.
In Sec.~\ref{comp} we deal with a comparison of the theoretical results with representative experiments.
The paper concludes with a summary and an outlook to future work.
%

%
%%% Theory
\section{\label{theory} Patterns driven by combined AC and DC electric fields}
In a first systematic study on patterns in nematics driven by superimposed electric fields two square-form ac-voltages with frequencies $\omega_1 < \omega_c$ and $\omega_2 > \omega_c$ have been considered \cite{John:2004}.
Later on, in more recent experiments \cite{Pietsch:2010} the case of two superimposed harmonic voltages of the form
\begin{eqnarray}
\label{eq:super}
U = U_{low} \sin{\omega_1 t} + U_{high} \sin(\omega_2 t + \beta)
\end{eqnarray}
has been explored.
One finds in the $U_{high}-U_{low}$ plane a pattern-free region, which is simply connected.
The detailed shape of that region depends in a complicated manner on the choice of the two frequencies $\omega_1$, $\omega_2$ and also on the phase shift $\beta$.
An exact theoretical analysis of the various scenarios is missing so far and would be quite demanding within the framework of the SM.
In order to simplify the situation by reducing the number of parameters in Eq.~(\ref{eq:super}) we have considered the case $\omega_1 =0$, i.e., a superposition of a dc- and an ac-voltage of the following form:
\begin{eqnarray}
\label{eq:superacdc}
U = U_{dc} + \sqrt{2} U_{ac} \cos(\omega t) \,.
\end{eqnarray}
We will consider only moderate ac-frequencies $\omega \tau_q >0.1$ as in  most experiments in the past.
In this way we avoid the problematic region $\omega \rightarrow 0$ for EC.
On the other hand it opens the possibility to study the mutual interaction between flexodomains which exist for $U_{ac} =0$ and EC patterns for $U_{dc} =0$.
Furthermore we expect that a comparison with experiments would yield valuable information on possible modifications of the externally applied voltage inside the cell, which might happen in particular through the dc-component of the applied voltage, for instance by an accumulation of the mobile ions at the electrodes (Debye layers).
To analyze this situation we had to modify the SM code used in \cite{Nonsta:2008} by including the additional dc-voltage.
As a result the special reflection-time-shift symmetry of the SM alluded to above is broken.
For $\omega < \omega_c$ the time average of the director component $n_z$ remains finite and we use further the short-hand notion ``conductive'' to describe the resulting low-$q_c$ patterns.
For $\omega > \omega_c$ this pattern type is expected as well when $U_{dc} \gg U_{ac}$ in Eq.~(\ref{eq:superacdc}).
In contrast, for $U_{dc} \ll U_{ac}$, where $n_z$ as well as the pattern amplitude are time periodic, the high-$q_c$ patterns are denoted as ``dielectric''.
Instead of presenting extended parameter studies we will discuss some characteristic examples of the onset of EC due to the applied voltage of the form given in Eq.~(\ref{eq:superacdc}).
They have been obtained by a linear stability analysis of our SM code,  where a ``standard'' material parameter set of the nematic Phase~5 has been used \cite{TreiberWM:1997,Toth:2008}:
$\epsilon_{\perp}=5.25$,  $\epsilon_a=-0.184$,
$\sigma_{\perp}=10^{-8}$~$[\Omega$~m$]^{-1}$,
$\sigma_a/\sigma_{\perp}=0.7$; 
elastic constants in units of $10^{-12}$~N: 
$k_{11}=9.8$, $k_{22}=4.6$, $k_{33}=12.7$; 
viscosity coefficients in units of $10^{-3}$~Pa~s: 
$\alpha_1=-39$, $\alpha_2=-109.3$, $\alpha_3=1.5$, 
$\alpha_4=56.3$, $\alpha_5=82.9$, $\alpha_6=-24.9$.
%

%%%
\subsection{\label{linstab} Numerical stability
analysis}
\begin{figure}[ht]
\centering
\includegraphics[width=7cm]{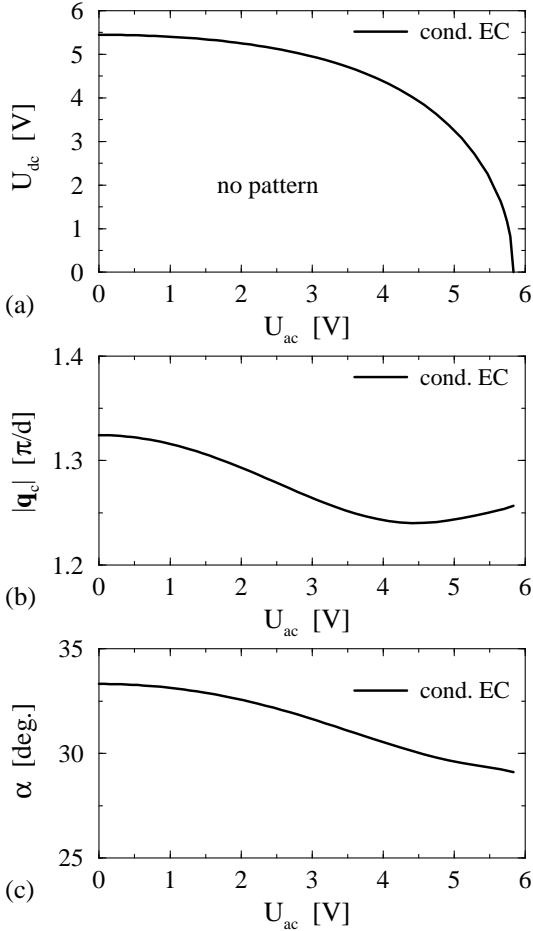}
\caption{\label{fig:Ucondcond} 
Phase diagram for EC patterns in the $U_{dc}-U_{ac}$ plane under combined dc- and ac-voltages with $\omega \tau_q =0.3$ ($f =10$~Hz) in the conductive regime:
Boundary curve enclosing the pattern-free region (a); the critical wave number $|\bm q_c|$ (b); angle $\alpha$ between the critical wave vector $\bm q_c$ and the $x$-axis (c) along the boundary curve.
Material parameters of Phase~5, thickness $d=10$~$\mu$m, flexocoefficients
$e_1=e_3=0$.} 
\end{figure}

Let us start with the case of an ac-frequency $\omega \tau_q =0.3$ which leads for $U_{dc} =0$ to EC rolls of the conductive symmetry.
The resulting quarter-ellipsoidal pattern-free region in Fig.~\ref{fig:Ucondcond}(a) is limited by a smooth curve.
With increasing $U_{ac}$ the critical value of $U_{dc}$ decreases until one finds for $U_{dc} =0$ the critical value of pure ac-voltage driven conductive rolls.
Moving along the boundary curve, starting from the left where $U_{ac} =0$ the EC roll patterns remain practically stationary; their wave numbers $|\bm q_c|$ and angles $\alpha$ between $\bm q_c$ and the preferred $x$-direction slightly decrease [see Figs.~\ref{fig:Ucondcond}(b), (c)].
%

%
% Figure 1
%
%

%
% Figure 2
%
\begin{figure}[ht]
\centering
\includegraphics[width=7cm]{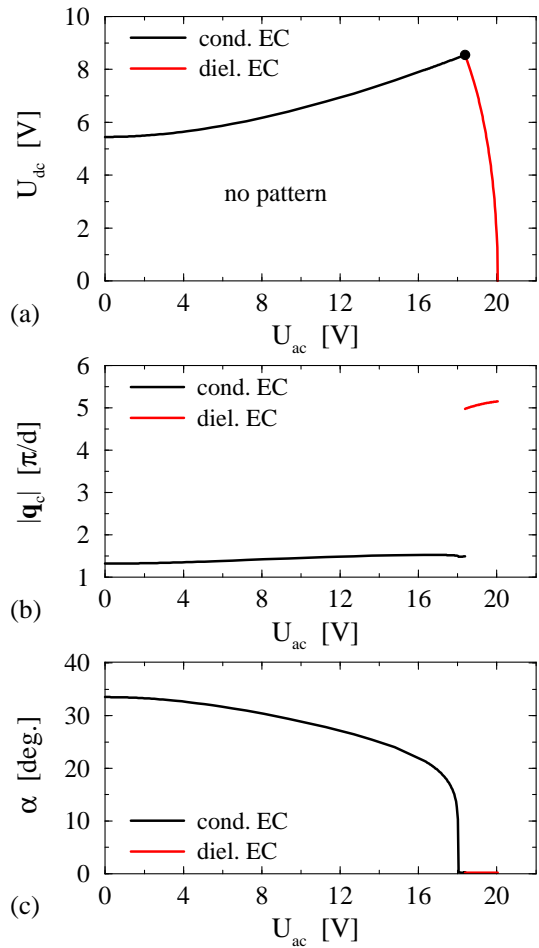}
\caption{\label{fig:Uconddiel} 
(Color online) Phase diagram for EC patterns in analogy to Fig.~\ref{fig:Ucondcond} except the ac-voltage component with $\omega \tau_q =6$ ($f =200$~Hz) in the dielectric regime.
The crossover between the conductive EC patterns and the dielectric ones is marked by the filled circle in (a).}
\end{figure}
Next we consider the situation with ac-frequencies above $\omega_c \tau_q \approx 2$ (see Fig.~\ref{fig:Uconddiel}).
The threshold at $U_{ac} =0$ is given by the threshold for pure dc-driving and is thus the same as in Fig.~\ref{fig:Ucondcond}(a).
With increasing $U_{ac}$ the pattern remains of conductive type along the upper boundary but, in contrast to the case of small $\omega$ [see Fig.~\ref{fig:Ucondcond}(a)], the critical value of $U_{dc}$ increases.
Starting alternatively at the threshold of dielectric rolls at $U_{dc} =0$ and increasing $U_{dc}$ the pattern of dielectric type persists along the right boundary curve in Fig.~\ref{fig:Uconddiel}(a).
As to be expected, these two boundary curves associated to patterns of different type and of substantially different wave numbers [Fig.~\ref{fig:Uconddiel}(b)] are not joining smoothly.
Note that similar smooth and non-smooth boundary curves have been observed before in the case of two superimposed harmonic voltages as well \cite{Pietsch:2010}.
%

%
% Figure 3
%
\begin{figure}[ht]
\centering
\includegraphics[width=7cm]{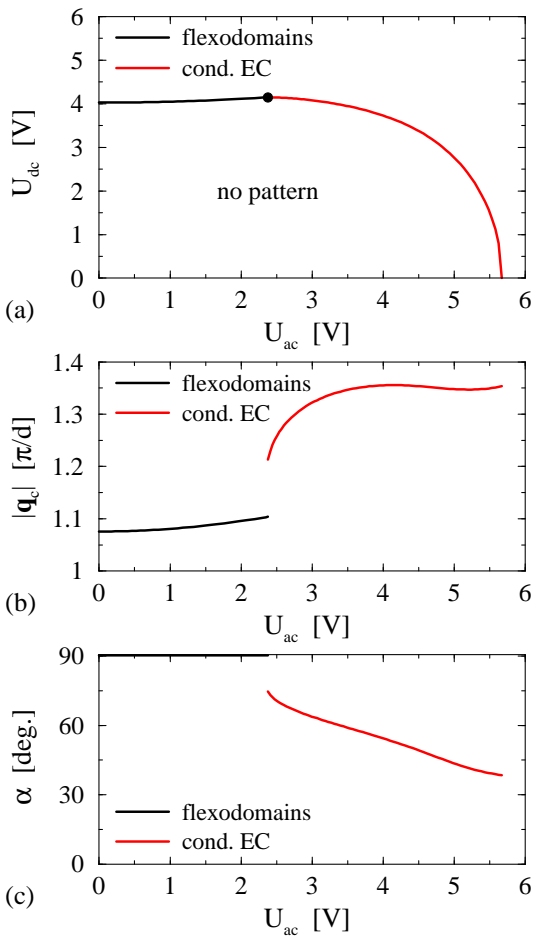}
\caption{\label{fig:flexcond} 
(Color online) Phase diagram for flexodomains and conductive EC rolls for the ac-voltage component with $\omega \tau_q =0.3$ ($f =10$~Hz) in analogy to Fig.~\ref{fig:Ucondcond} except for finite flexocoefficients $e_1 =12$~pC/m, $e_3 =0$.
The crossover between flexodomains and conductive EC rolls is marked by the filled circle in (a).}
\end{figure}
%

%
% Figure 4
%
\begin{figure}[ht]
\centering
\includegraphics[width=7cm]{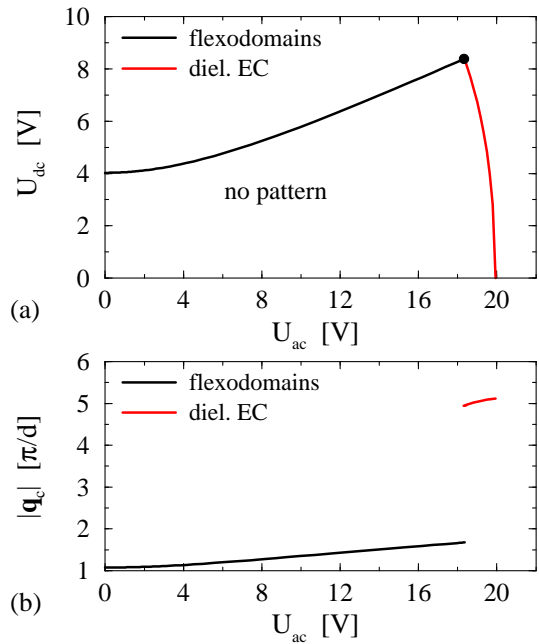}
\caption{\label{fig:flexdiel} 
(Color online) Phase diagram for flexodomains and dielectric EC rolls in analogy to Fig.~\ref{fig:flexcond}, but for the much higher ac-frequency $\omega \tau_q =6$ ($f =200$~Hz).
The crossover between flexodomains and dielectric EC rolls is marked by the filled circle in (a).}
\end{figure}
Finally we consider the competition between flexo\-domains and EC pattern, either in the conductive or in the dielectric regime. 
For that purpose we have to modify the Phase~5 material parameter set listed before, for which the bifurcation to EC prevails for all $\omega \tau_q$.
One has to include flexoelectricity into the SM parameterized by the flexocoefficients $e_1$, $e_3$.
The threshold voltage for flexodomains, $U_c^{flex}$, at $\omega =0$ is proportional to $|e_1 - e_3|$ (see, e.g., Ref.~\cite{Flexo:2011}) and by choosing for simplicity $e_1 =12$~pC/m, $e_3 =0$ we obtain in fact a bifurcation to flexodomains at $\omega =0$.
As demonstrated in Figs.~\ref{fig:flexcond}(a) and \ref{fig:flexdiel}(a) we obtain in the context of this pure model study again pattern-free regions which look very similar to the ones described before in Figs.~\ref{fig:Ucondcond}, \ref{fig:Uconddiel} where only EC patterns are involved.
According to Fig.~\ref{fig:flexcond} for the ac-frequency $\omega \tau_q = 0.3$ in the conductive regime of EC we find first flexodomains, when following the upper transition curve with increasing $U_{ac}$.
As characteristic for flexodomains, the angle $\alpha$ between the critical wave vector $\bm q_c$ and the $x-$axis remains at $\alpha =90^{\circ}$, while $|\bm q_c|$ slowly increases.
At about $U_{ac} \approx 2.5$~V a crossover to EC patterns takes place and the transition curve in Fig.~\ref{fig:flexcond}(a) monotonically bends down when further increasing $U_{ac}$.
Furthermore one observes jumps in the wave number $|\bm q_c|$ and in the angle $\alpha$ at the crossover point [Figs.~\ref{fig:flexcond}(b), (c)].
While the latter quantity decreases substantially with increasing $U_{ac}$, the former one increases.
The corresponding scenario for the ac-driving in the dielectric regime ($\omega \tau_q = 6$) is presented in Fig.~\ref{fig:flexdiel} and looks qualitatively similar.
With increasing ac-voltage at about $U_{ac} \approx 18$~V we find the transition from flexodomains to dielectric EC rolls, which are confined to a much smaller $U_{ac}$ interval compared to Fig.~\ref{fig:flexcond}.
In contrast to Fig.~\ref{fig:flexcond}(a) at the crossover point the slope of the transition curve dramatically changes and the jump of $|\bm q_c|$ is much larger.
The angle $\alpha$ between the critical wave vector $\bm q_c$ and the $x-$axis switches from $\alpha =90^{\circ}$ to $\alpha =0$ (not shown).
%

%%%
\subsection{\label{qual} Qualitative analysis}
In the following we give some rough arguments to support the exact numerical results on the general shape of the pattern-free regions shown in the figures above.
In principle we go back to the basic elements of the positive feedback mechanism, named commonly after Carr and Helfrich \cite{Carr:1969,Helfrich:1969}, which governs the EC instability.
In essence the reasoning is as follows: Any out-of-plane distortion of the originally planar director configuration (finite $n_z$) leads to charge separation and to a finite charge density $\rho_{el}$.
As a consequence, in the presence of an electric field of strength $E_z$ across the nematic layer the Coulomb force $\rho_{el} E_z$ appears in the velocity equation, by which a velocity field with the component $v_z$ is excited.
A necessary condition for destabilization of the basic planar state towards EC patterns is the reinforcement of the original director distortion by the resulting viscous torque $\propto v_z$ on the director.
Details are discussed for instance in Ref.~\cite{Nonsta:2008}.
In particular the time symmetry of the patterns plays an important role.
In the case of a dc-voltage or a low-$\omega$ ac-voltage (conductive regime) both $n_z$ and $v_z \propto \rho_{el} E_z$ are virtually time independent, while both $E_z$ and $\rho_{el}$ oscillate sinusoidally for small $\omega$.
For large $\omega$ the time symmetry is reversed.
Thus it is clear from the beginning that the time-independent torques originating from a pure dc-electric field and an ac-electric field in the conductive regime allow for their optimal cooperation.
The following paragraphs are devoted to a more detailed analysis on the basis of a perturbative treatment of the combined action of ac- and dc-voltages.
We will focus on two particular sections of the boundary curves of the various pattern-free regions discussed in Sec.~\ref{linstab}, namely on small ac perturbations of the purely dc-voltage driven EC patterns and alternatively on small dc perturbations of purely ac-voltage driven EC patterns.
The appropriate lowest-order ansatz for the out-of-plane component of the director, which vanishes at the upper and the lower horizontal boundary plates (at $z = \pm d/2$) is given as: 
\begin{eqnarray}
\label{eq:nz_ansatz}
n_z(x,y,z,t) = N_z(t) \cos(\pi z / d) \cos(q x + p y) \,,
\end{eqnarray}
with $\bm q_c = (q, p)$.
The amplitude $N_z$ is determined by the time symmetry of the EC modes.
At finite $n_z$ the dielectric displacement $\bm D$ leads via $\rho_{el} = \nabla \cdot \bm D$ to the charge density $\rho_{el}$ in the form
\begin{eqnarray}
\label{eq:rho_el_ansatz}
\rho_{el}(x,y,z,t) = \bar{\rho}_{el}(t) \cos(\pi z / d) \sin(q x + p y) \,.
\end{eqnarray}
Let us start with the case $U_{dc} \gg U_{ac}$ in Eq.~(\ref{eq:superacdc}).
Concentrating first on a small ac frequency $\omega$ the $z$-component of the applied electric field can be represented as:
\begin{eqnarray}
\label{eq:E0}
E_z = E_{dc} + \delta E_{ac} \cos(\omega t) \,,
\end{eqnarray}
where $E_{dc}$ corresponds to the critical dc-field and $\delta E_{ac} \cos(\omega t)$ describes the appropriate time-periodic perturbation in the conductive regime ($\omega \tau_q \ll1$), where on the other hand both the director field and the velocity are time independent in leading order.
Consequently we use the following ansatz for the amplitude $N_z$ of the out-of-plane director component:
\begin{eqnarray}
\label{eq:nz_dc_ac_cond}
N_z = N_z^{dc} + \delta N_z^{ac} \,,
\end{eqnarray}
where $N_z^{dc}, \delta N_z^{ac} >0$.
Using Eqs.~(\ref{eq:nz_ansatz}) and (\ref{eq:nz_dc_ac_cond}) the corresponding amplitude of the charge density $\bar{\rho}_{el}$ can be calculated (for details see, e.g., Eq.~(A11) in Ref.~\cite{Nonsta:2008}).
For rolls driven by dc-field only the stationary correction  of the $z$-component of the Coulomb force ($\propto \bar{\rho}_{el} E_z$) in the Navier-Stokes equation is relevant.
In leading order in the perturbations $\delta N_z^{ac}$, $\delta E_{ac}$ one finds
\begin{eqnarray}
\label{eq:fb_dc_ac_cond}
\bar{\rho}_{el} E_z = E_{dc}^2 (N_z^{dc} + \delta N_z^{ac}) \rho_0 \,,
\end{eqnarray}
where $\rho_0$ is given by
\begin{eqnarray}
\label{eq:rho_0}
\rho_0 =
\frac{q \epsilon_0 \epsilon_{\perp} (\sigma_a/\sigma_{\perp} - \epsilon_a/\epsilon_{\perp}) }
{1 + (\sigma_a/\sigma_{\perp}) q^2 /(q^2+p^2+\pi^2/d^2)} \,.
\end{eqnarray}
For $(\sigma_a/\sigma_{\perp} - \epsilon_a/\epsilon_{\perp}) >0$ which holds for Phase~5, the positive correction $E_{dc}^2 \delta N_z^{ac} \rho_0$ of Coulomb force leads to an enhancement of $v_z$ and thus of the destabilizing viscous torque.
Consequently, the EC instability occurs at a reduced value of $E_{dc}$, in line with the bending down of the upper boundary curve in Fig.~\ref{fig:Ucondcond}(a).
In the following we continue with the case $U_{dc} \gg U_{ac}$ but consider a high-frequency ac-voltage perturbation in the dielectric regime.
In analogy to Eq.~(\ref{eq:nz_dc_ac_cond}) the appropriate lowest-mode ansatz involves now the generic time-periodic perturbation in the dielectric regime as follows:
\begin{eqnarray}
\label{eq:nz_dc_ac_diel}
N_z = N_z^{dc} + \delta N_z^{ac} \cos(\omega t) \,.
\end{eqnarray}
Following the same strategy as before we arrive after some algebra at:
\begin{eqnarray}
\label{eq:fb_dc_ac_diel}
\bar{\rho}_{el} E_z = (E_{dc}^2 N_z^{dc}
- \frac{1}{2} E_{dc} \delta E_{ac} \delta N_z^{ac}) \rho_0 \,,
\end{eqnarray}
The main difference to the low $\omega$ case considered before is, that the Coulomb force driving the pure $U_{dc}$ patterns (i.e., $E_{dc}^2 N_z^{dc} \rho_0$) acquires a negative correction which depends quadratically on the ac-perturbations.
Accordingly, the threshold $E_{dc}$ has to increase in line with Fig.~\ref{fig:Uconddiel}(a).
The following considerations deal with the second main part, $U_{ac} \gg U_{dc}$, of our perturbative analysis regarding the effects of combined ac- and dc-voltages.
In analogy to Eq.~(\ref{eq:E0}) we use the following ansatz for the electric field:
\begin{eqnarray}
\label{eq:E0_ac_small_dc}
E_z = E_{ac} \cos(\omega t) + \delta E_{dc} \,,
\end{eqnarray}
where $E_{ac}$ corresponds to the critical ac-field and $\delta E_{dc}$ is a small dc-perturbation.
First consider the case of an ac-field with the frequency $\omega \tau_q \ll 1$ in the conductive regime.
According to the time-symmetries of the linear unstable modes either driven by an ac-voltage in the conductive regime and or by a dc-voltage, one uses for the amplitude of the out-of-plane director component the following ansatz:
\begin{eqnarray}
\label{eq:nz_ac_cond_dc}
N_z = N_z^{ac} + \delta N_z^{dc} \,,
\end{eqnarray}
where the first term corresponds to the dominant ac-field driven director distortion and the second term appears due to the dc-field perturbation.
Calculating the electric charge density as before we arrive finally in the leading order with respect to the perturbations $\delta N_z^{ac}$, $\delta E_{ac}$ and for $\omega \tau_q \ll 1$ at
\begin{eqnarray}
\label{eq:fb_ac_cond_dc}
\bar{\rho}_{el} E_z = \frac{1}{2} E_{ac}^2 (N_z^{ac} + \delta N_z^{dc}) \rho_0 \,.
\end{eqnarray}
Obviously the dc electric field perturbation causes an increase of the Coulomb force and thus of $v_z$ as well, resulting in an enhanced destabilizing viscous torque.
Consequently the critical $E_{ac}$ is decreasing in the presence of $\delta E_{dc}$ and the vertical threshold curve must bend to the left in line with Fig.~\ref{fig:Ucondcond}(a).
Finally we arrive again at the case $U_{ac} \gg U_{dc}$ but consider a high-frequency ac-voltage with the frequency $\omega \tau_q \gg 1$ in the dielectric regime.
The appropriate ansatz for the lowest-mode director component, $N_z$,  reads as follows:
\begin{eqnarray}
\label{eq:nz_ac_diel_dc}
N_z = N_z^{ac} \cos(\omega t) + \delta N_z^{dc} \,.
\end{eqnarray}
The rolls with dielectric symmetry are driven by the oscillatory component of the Coulomb force $\propto \cos(\omega t)$, which in leading order in the perturbations $\delta N_z^{ac}$, $\delta E_{ac}$ and for $\omega \tau_q \gg 1$ is given by:
\begin{eqnarray}
\label{eq:fb_ac_diel_dc}
\bar{\rho}_{el} E_z = ( \frac{1}{2} E_{ac}^2 N_z^{ac}
+ E_{ac} \delta E_{dc} \delta N_z^{dc}) \rho_0 \cos(\omega t) \,.
\end{eqnarray}
The correction to the pure ac-field is now of the second order in the perturbations $\delta N_z^{dc}$, $\delta E_{dc}$ and leads to an increase of the Coulomb force.
Accordingly, the critical value of $E_{ac}$ has to decrease in line with the bending to the left of the right boundary curve in Fig.~\ref{fig:Uconddiel}(a).
There exists no obvious way to capture qualitatively the competition between flexodomains and EC patterns, as demonstrated in Figs.~\ref{fig:flexcond}, \ref{fig:flexdiel}.
For instance in the flexodomains with $\bm q_c \parallel \bm{\hat y}$ a finite charge density, which is responsible for the EC instability and has played a crucial role in our considerations before, does not exist.
%

%
%%% Comparison with experiments
\section{\label{comp} Comparison with experiments}
\begin{figure}[ht]
\centering
\includegraphics[width=7cm]{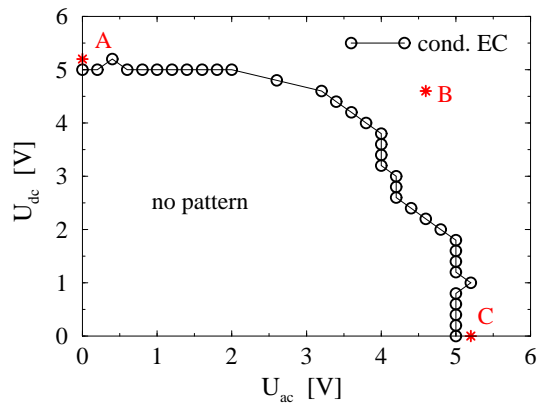}
\caption{\label{fig:expcond} 
Experimental boundary curve of the pattern-free region in the $U_{dc}-U_{ac}$ plane in the presence of combined dc- and ac-voltages for $f =10$~Hz, in the conductive regime for Phase~5.
Cell thickness $d=11.4$~$\mu$m.
Stars indicate the locations where the snapshots of Fig.~\ref{fig:qccond} were taken.}
\end{figure}

\begin{figure}[ht]
\centering
(A)\includegraphics[width=2.3cm]{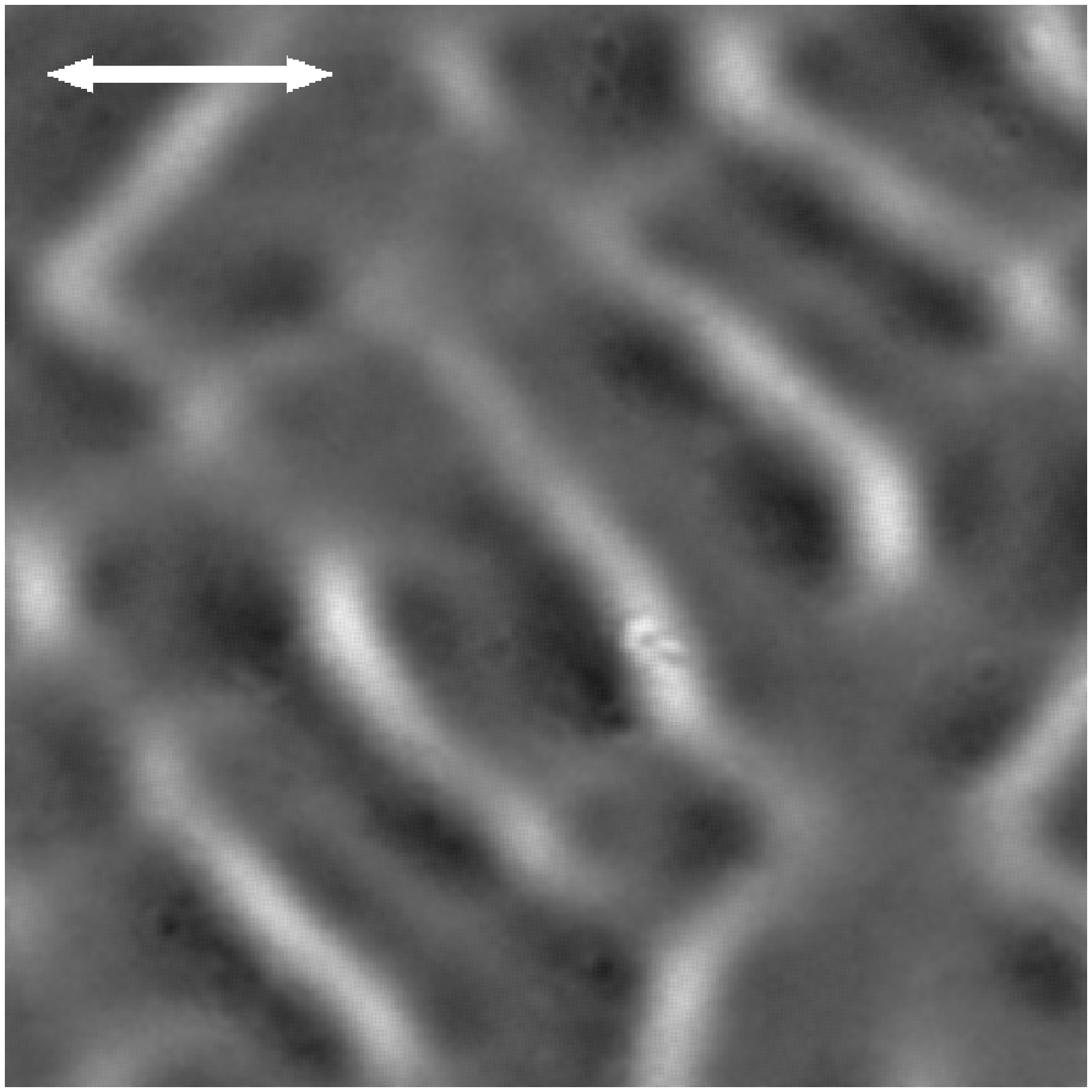}
(B)\includegraphics[width=2.3cm]{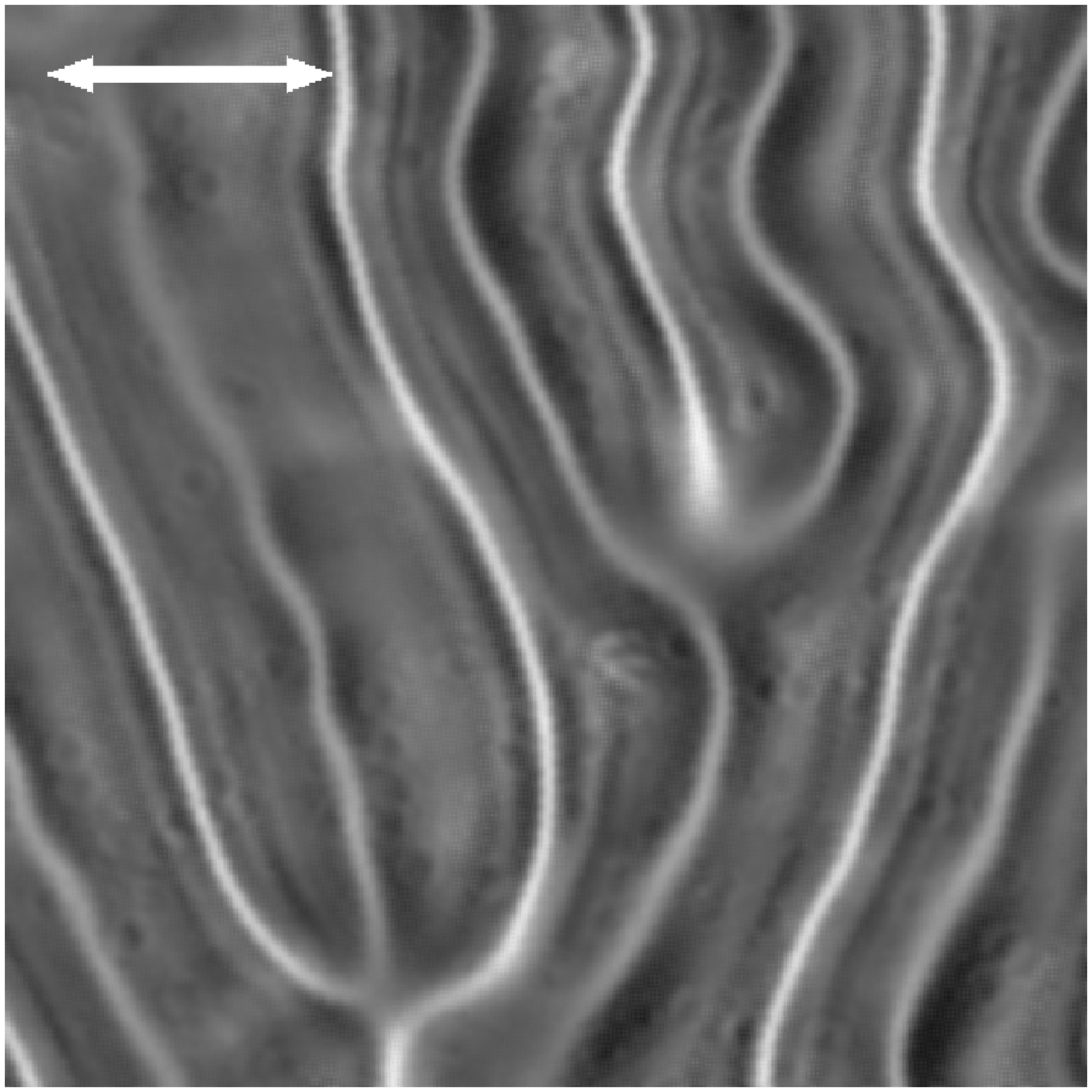}
(C)\includegraphics[width=2.3cm]{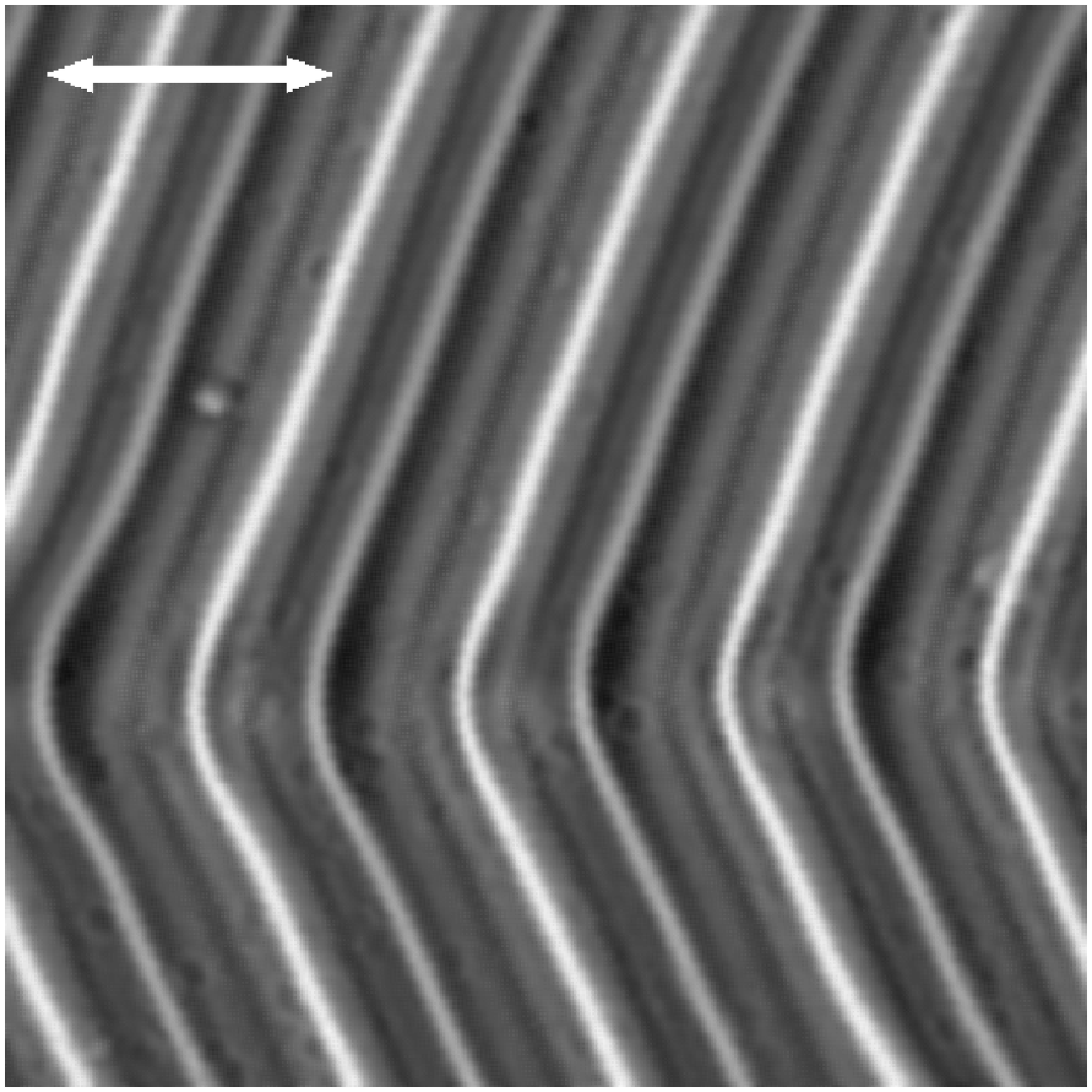}
\caption{\label{fig:qccond} 
Snapshots of EC patterns along the boundary curve at the three locations indicated in Fig.~\ref{fig:expcond}: (A) pure dc-driving; (B) superimposed  dc- and ac-voltages; and (C) pure ac-driving.
The wave number $q_c$ and the obliqueness angle $\alpha$ of the observed oblique conductive rolls are: 
(A) $q_c \approx 1.42 \pi/d $, $\alpha \approx \pm 37^{\circ}$; 
(B) $q_c \approx 1.12 \pi/d $, $\alpha \approx \pm 26^{\circ}$; 
(C) $q_c \approx 1.30 \pi/d $, $\alpha \approx \pm 22^{\circ}$.
The arrow bars of length $20$~$\mu$m are parallel to the initial planar director orientation $\bm n \parallel \bm{\hat x}$.}
\end{figure}

In the following section we compare the theoretical calculations with experiments carried out on the nematic Phase~5.
In Fig.~\ref{fig:expcond} we show an example of the EC pattern-free region in the presence of combined dc- and ac-voltages at fairly low ac-frequency ($f=10$~Hz), much below the crossover $f_c$, thus corresponding to the conductive regime of EC.
The shape of that region matches satisfactorily the theoretical phase diagram shown in Fig.~\ref{fig:Ucondcond}(a).
In figure~\ref{fig:qccond} one finds representative sections of shadowgraph images taken slightly above onset, at the locations marked by stars in Fig.~\ref{fig:expcond}.
They are indeed of the conductive type.
Experimental oblique roll patterns are rarely perfect.
Apart from the appearance of dislocations one observes so-called ``zig'' and ``zag'' patches with the symmetry degenerated wave vectors $\bm q_c =(q,\pm p)$, which are separated by grain boundaries.
Nevertheless, we were able to extract via a Fourier analysis the experimental values of $\bm q_c$.
The obliqueness angle $\alpha$ between the wave vector $\bm q_c$ and the $x$-axis is gradually decreasing from $\alpha \approx 37^{\circ}$ to $\alpha \approx 22^{\circ}$ as $U_{ac}$ increases.
The wave number acquires its maximum $q_c \approx 1.42 \pi/d$ for $U_{ac} =0$, passes then through a minimum ($q_c \approx 1.12 \pi/d$) when $U_{dc}$ and $U_{ac}$ are roughly comparable and rises again to the value $q_c \approx 1.3 \pi/d$ when $U_{dc} \rightarrow 0$.
The general change of $\bm q_c$ when moving from left to right along the limiting curve of the pattern-free region in Fig.~\ref{fig:expcond} is in satisfactory agreement with the theoretical results shown in Fig.~\ref{fig:Ucondcond}.
%

%
% Figure 5
%
%

%
%
% Figure 6
%
\begin{figure}[ht]
\centering
(a)\includegraphics[width=7cm]{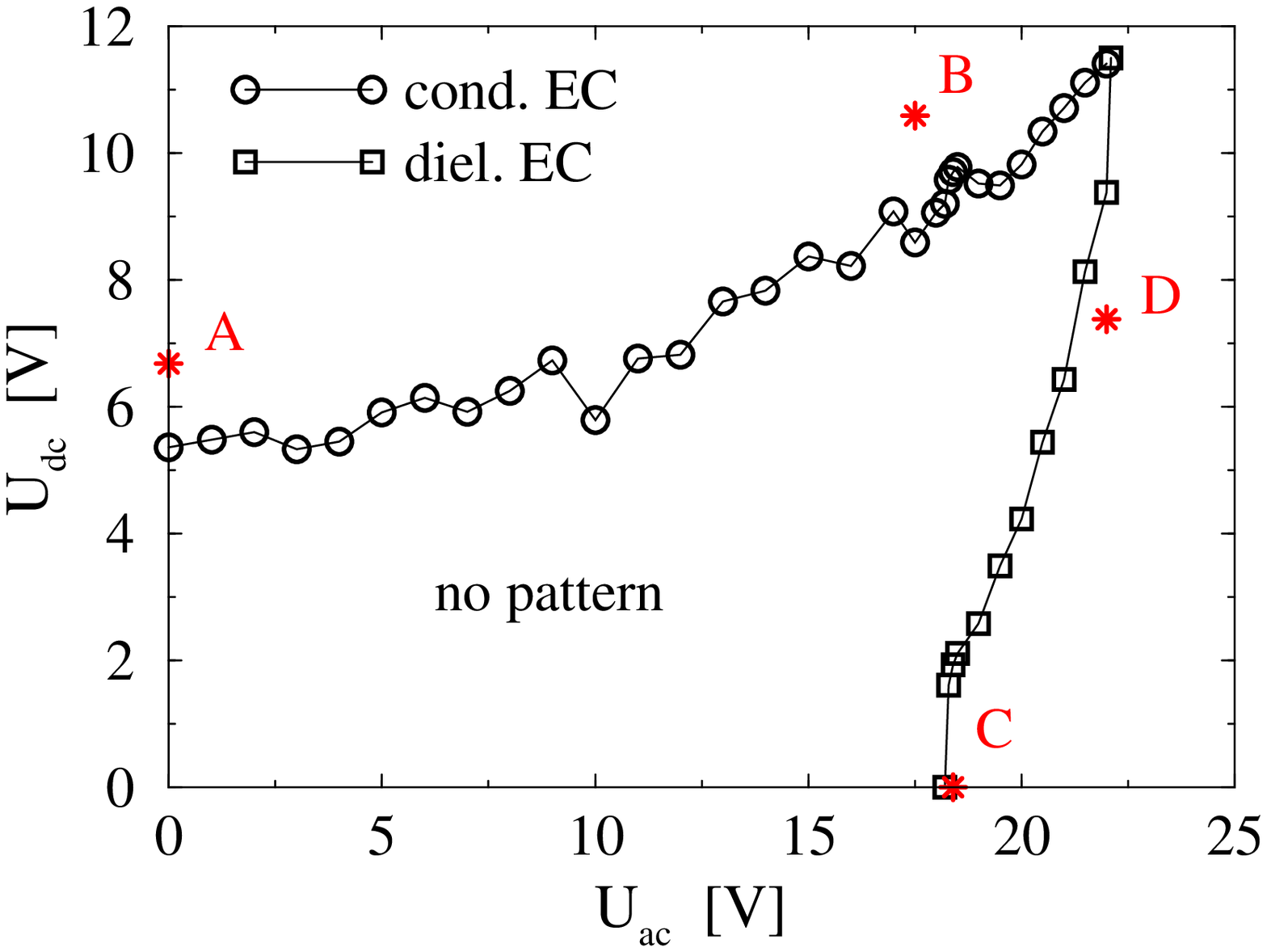}
(b)\includegraphics[width=7cm]{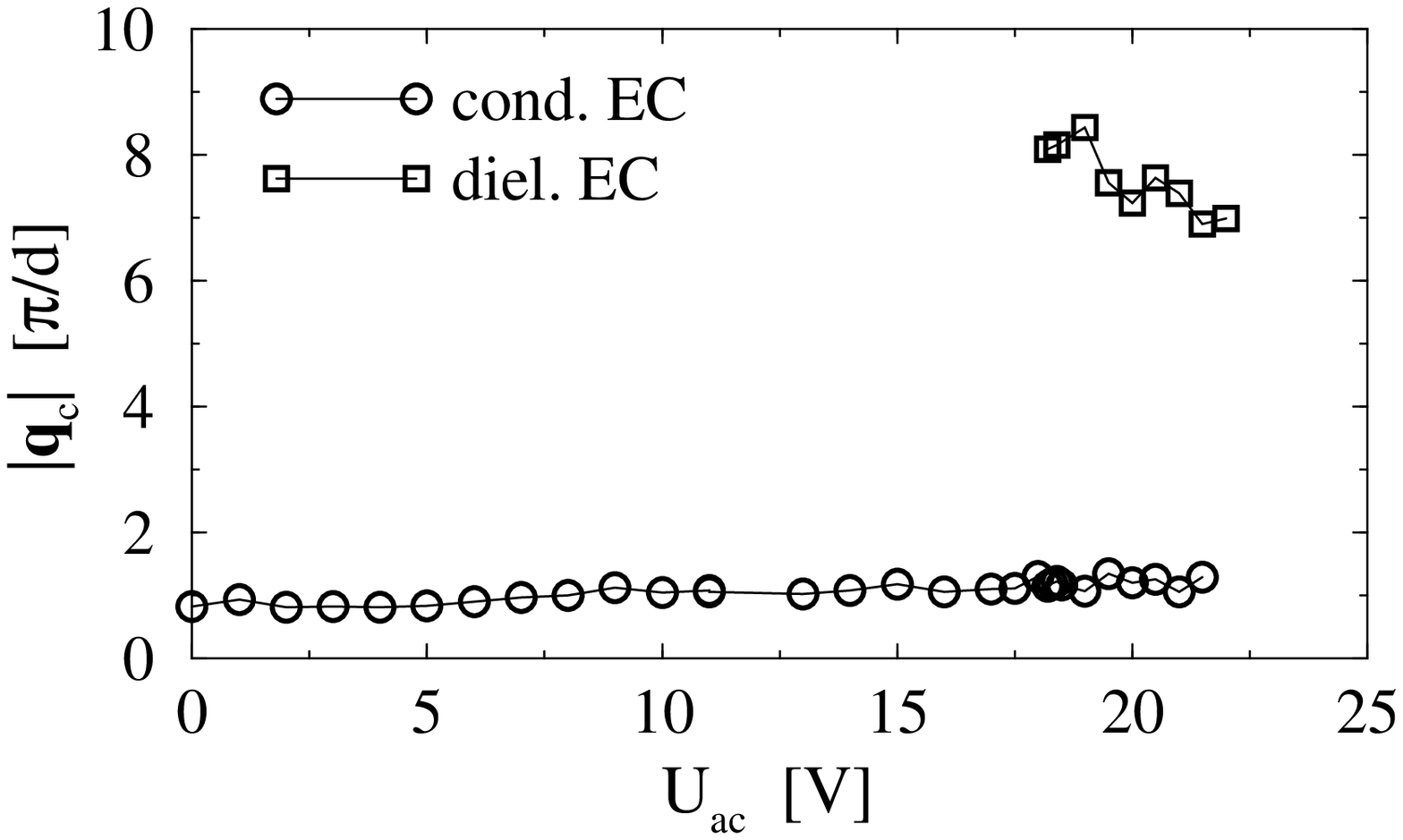}
\caption{\label{fig:expdiel}
Experimental phase diagram of EC patterns for Phase~5 in the $U_{dc}-U_{ac}$ plane for an ac-frequency $f =200$~Hz, in the dielectric regime: Boundary curve enclosing the pattern-free region (a); the critical wave number $|\bm q_c|$ along the boundary curve (b).
Cell thickness $d=10.8$~$\mu$m.
Stars indicate the locations where the snapshots of Fig.~\ref{fig:expdielpi} were taken.}
\end{figure}

\begin{figure}[ht]
\medskip
\medskip
\centering
(A)\includegraphics[width=2.3cm]{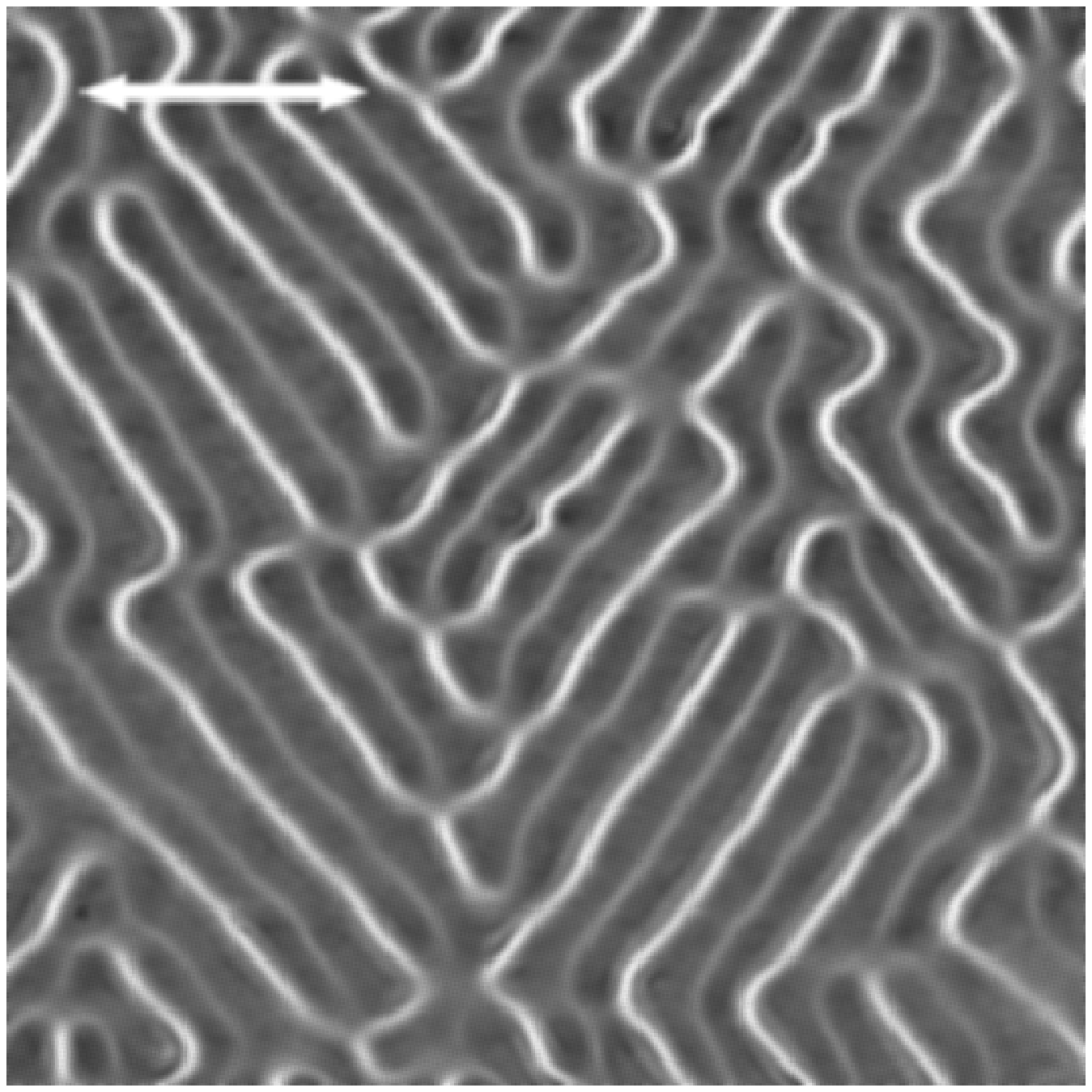}
(B)\includegraphics[width=2.3cm]{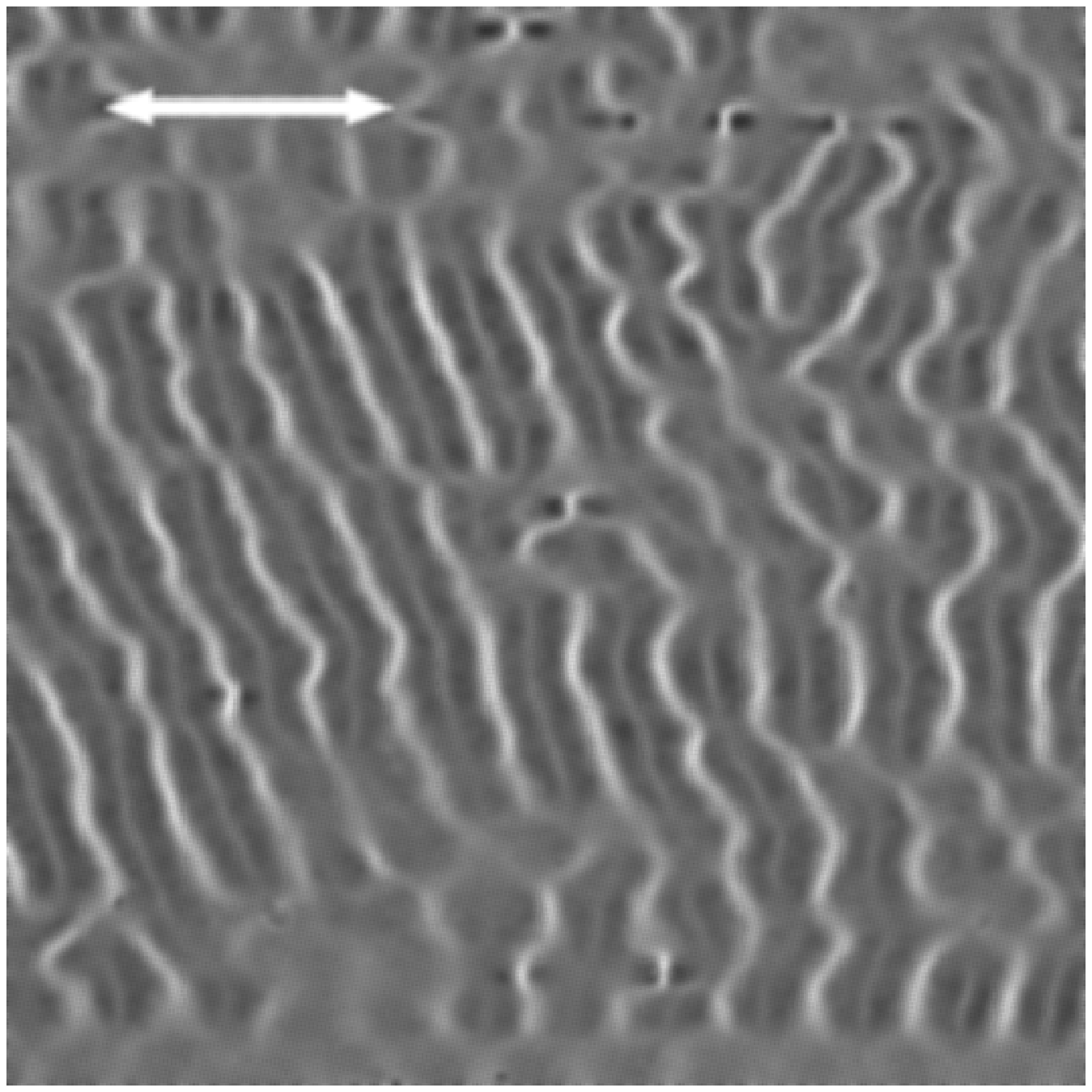}
\\
\medskip
\medskip
(C)\includegraphics[width=2.3cm]{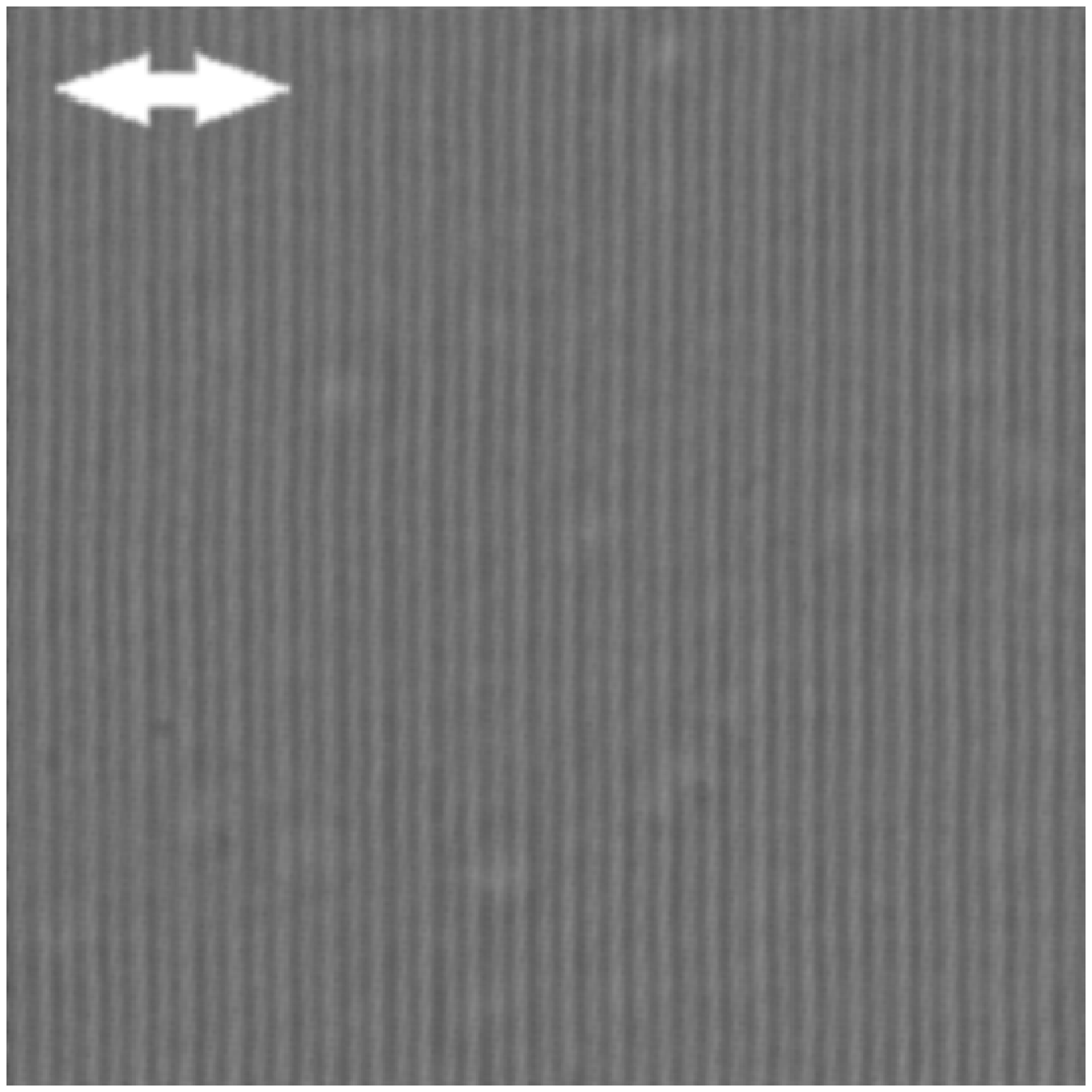}
(D)\includegraphics[width=2.3cm]{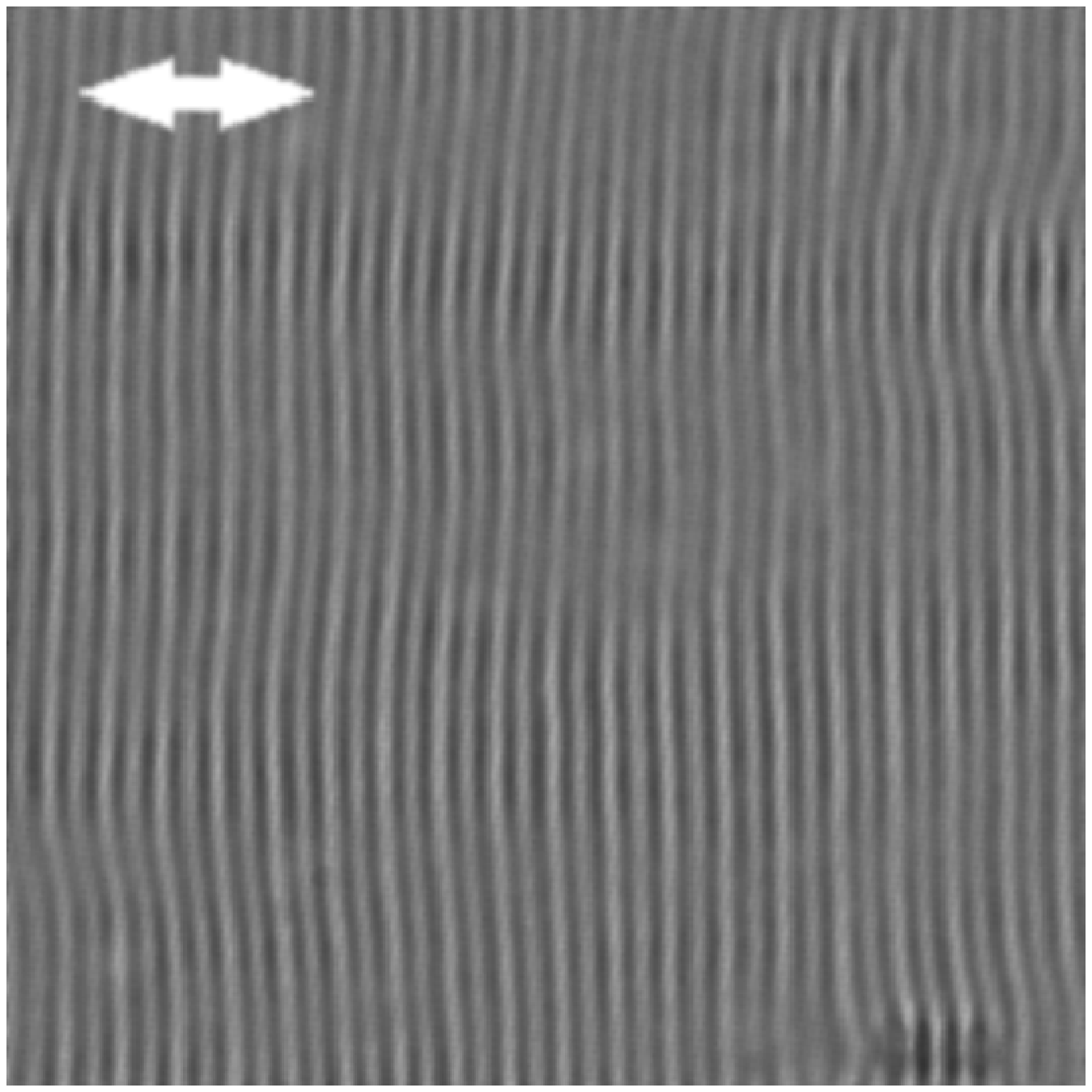}
\caption{\label{fig:expdielpi} 
Snapshots of EC patterns along the boundary curve at the four locations indicated in Fig.~\ref{fig:expdiel}:
(A) oblique rolls at pure dc-driving; (B) oblique rolls at superposed dc- and ac-voltages; (C) dielectric rolls at pure ac-driving; and (D) dielectric rolls at superposed dc- and ac-driving.
The arrow bars [of length $50$~$\mu$m for (A), (B), and $20$~$\mu$m for (C), (D)] are parallel to the initial planar director orientation $\bm n \parallel \bm{\hat x}$.}
\end{figure}

As an example for the combination of a dc-voltage with an ac-voltage in the dielectric regime, we present in Fig.~\ref{fig:expdiel}(a) the phase diagram for a large ac-frequency ($f=200$~Hz, above the crossover $f_c$), where a competition occurs between conductive (low $q_c$) and dielectric (high $q_c$) rolls.
In agreement with the theoretical shape of the pattern-free region shown in Fig.~\ref{fig:Uconddiel}(a), the upper and the right boundary lines do not merge smoothly.
Instead, they cross at a finite angle, indicating the sharp morphological transition between conductive and dielectric EC rolls.
The corresponding jump in the critical wave number $|\bm q_c|$, shown in Fig.~\ref{fig:expdiel}(b), compares well with Fig.~\ref{fig:Uconddiel}(b).
The qualitative change of the pattern type along the transition curve is clearly documented in the experiments: Figure~\ref{fig:expdielpi}(A)-(B) shows representative snapshots of conductive patterns for smaller $U_{ac}$ before they are replaced by dielectric ones at larger $U_{ac}$ [see Fig.~\ref{fig:expdielpi}(C)-(D)].
A closer look at Figs.~\ref{fig:Uconddiel}(a), \ref{fig:expdiel}(a) reveals, however, a certain discrepancy between theory and experiment.
While the theoretical as well as the experimental upper transition lines in the conductive regime considerably curve upwards, the experimental right transition line in the dielectric regime bends to the right and not to the left as in the theory.
%

%
% Figure 7
%
%

%
% Figure 8
%
%

%
%%% Summary and conclusions
\section{\label{summary} Summary and outlook}
It is very satisfactory that in general our theoretical results are well confirmed by the experiments.
In particular it is reassuring that the measured (external) voltages and critical wave vectors compare well with the theoretical (internal) voltages obtained by the proper choice of the conductivity.
One is accustomed to such a good agreement in conventional EC experiments with pure ac-voltage driving.
As already mentioned at the end of Sec.~\ref{comp}, however, the boundary curves in the dielectric regime under combined ac- and dc-driving show an opposite curvature in theory and experiment.
This points to the fact that in the theory some mechanisms seem to be missing.
One of them comes immediately to mind.
Due to the applied dc-voltage a certain fraction of the mobile ions will certainly move to the electrodes.
As a consequence, the conductivity of the NLC decreases with increasing $U_{dc}$ which, typically, leads to modifications of the threshold voltage of EC.
Certainly more systematic, but also very time consuming, experiments are desirable in the future.
In any case, the ac-frequency $\omega$ and the cell thickness $d$ are  important parameters, which need a thorough exploration.
Furthermore one should use different nematic materials.
A possible candidate to test for instance the phase diagram in Fig.~\ref{fig:flexdiel} might be the nematic Phase~4 (with material parameters comparable to Phase~5), where already a transition between flexodomains under pure dc-driving to dielectric EC rolls under pure ac-driving at very low frequencies has been observed \cite{bib:May-2008}. 
From the theoretical point of view, certainly a more detailed analysis is needed as well.
It looks very plausible that the dc-voltage profile across the cell in the basic planar state will substantially deviate from linearity as assumed in the the SM.
For instance, the cell may consist of regions with strong field variations near the electrodes (over the Debye layers) and weaker ones in the central part of the cell.
One would expect that such field gradients have a larger influence on patterns with shorter wavelengths (i.e., in dielectric rolls), which are comparable to the characteristic length scales of the electric potential variations.
To account for such variations, the description of the NLC as an ohmic conductor has to be abandoned and replaced by a more detailed description of the mobile ions.
Major efforts have been made in the past to describe the ionic effects on the electrical conductivity of isotropic liquids (see, e.g., \cite{Castellanos:1998} and references therein).
An important role plays the formation of the so-called double layers at the electrodes, which depend on the detailed design of the electrodes.
One finds notions like ``blocking'', ``injecting'', ``charge carrier absorbing'' electrodes \cite{Turnbull:1973,Atten:1975}.
To include such mechanism into the description of electrically driven pattern forming instabilities in nematics the ``weak electrolyte model'' (WEM) has been formulated in the past \cite{Treiber:1995}.
So far it has only been used to explain the occurrence of traveling waves in certain EC experiments under pure ac-driving, which are excluded within the framework of the SM \cite{Treiber:1996,TreiberWM:1997}.
It is expected, however, that a more complete analysis of this model would give important insight into the complex physics of patterns developing under the combined action of ac- and dc electric fields.
%

%
% References

%
\end{document}